\documentclass[aps,prl,reprint,groupedaddress]{revtex4-1} 
\usepackage[dvipsnames]{xcolor}
\usepackage[colorlinks=true,urlcolor=blue,citecolor=purple,linkcolor=red]{hyperref}
\usepackage{amssymb,amsmath,amsfonts}
\usepackage{epsfig}
\usepackage{graphicx}
\usepackage{epstopdf}
\usepackage{natbib}

\def\clock{{\count0=\time
           \divide\count0 60
           \ifnum\count0<10 0\fi\the\count0
           \multiply\count0 -60 \advance\count0 \time
           :\ifnum\count0<10 0\fi \the\count0
         }}
\newcommand{\timestamp}{{\small\vbox{\hbox{\tt\jobname.tex}
\hbox{\the\day/\the\month/\the\year, \clock}}}}

\newcommand{\nn}{\nonumber}

\newcommand{\be}{\begin{eqnarray}}
\newcommand{\ee}{\end{eqnarray}}
\newcommand{\beq}{\begin{eqnarray}}
\newcommand{\eeq}{\end{eqnarray}}

\newcommand{\beqa}{\begin{eqnarray}}
\newcommand{\eeqa}{\end{eqnarray}}

\let\oldsqrt\sqrt
\def\sqrt{\mathpalette\DHLhksqrt}
\def\DHLhksqrt#1#2{%
\setbox0=\hbox{$#1\oldsqrt{#2\,}$}\dimen0=\ht0
\advance\dimen0-0.2\ht0
\setbox2=\hbox{\vrule height\ht0 depth -\dimen0}%
{\box0\lower0.4pt\box2}}

\begin{document}

\title{Action Principle for Newtonian Gravity}


\author{Dennis Hansen$^1$}
\author{Jelle Hartong$^2$}
\author{Niels A. Obers$^3$}

\email{dehansen@phys.ethz.ch}
\email{Jelle.Hartong@ed.ac.uk}
\email{obers@nbi.ku.dk}
\affiliation{$^1$ Institut f{\"u}r Theoretische Physik, Eidgen{\"o}ssische Technische Hochschule Z{\"u}rich\\ 
Wolfgang-Pauli-Strasse 27, 8093 Z{\"u}rich, Switzerland}
\affiliation{$^2$ School of Mathematics and Maxwell Institute for Mathematical Sciences,\\
University of Edinburgh, Peter Guthrie Tait road, Edinburgh EH9 3FD, UK } 
\affiliation{$^3$ The Niels Bohr Institute, Copenhagen University,\\
Blegdamsvej 17, DK-2100 Copenhagen \O , Denmark}



\begin{abstract}

We derive an action whose equations of motion contain the Poisson equation of Newtonian gravity. The construction requires a new notion of Newton--Cartan geometry based on an underlying symmetry algebra that differs from the usual Bargmann algebra. This geometry naturally arises in a covariant $1/c$ expansion of general relativity with $c$ being the speed of light.  By truncating this expansion at subleading order we obtain the field content and transformation rules of the fields that appear in the action of Newtonian gravity. The equations of motion generalize Newtonian gravity by allowing for the effect of gravitational time dilation due to strong gravitational fields. 

\end{abstract}



\maketitle




The idea that gravity is geometry was pioneered by Einstein in his celebrated theory of general relativity (GR). 
In GR, due to Einstein's equivalence principle, the underlying geometry is (pseudo-)Riemannian which ensures
that one has local Lorentz symmetry and hence the laws of physics locally reduce to those of special relativity.
However,  spacetime covariance is a property of any physical theory, which led Cartan \cite{Cartan1,Cartan2} 
(see also e.g.  \cite{Trautman63,Havas:1964zza}) to geometrize Newtonian gravity using what is known as Newton--Cartan (NC) geometry. 
The latter results from applying an equivalence principle that requires freely falling observers to see Galilean laws
of physics, giving rise to a geometry with local Galilean invariance. 

However, while the Poisson equation of Newtonian gravity can be geometrized using NC geometry, an outstanding question has been to find an action principle  for Newtonian gravity, paralleling the Einstein-Hilbert action in GR. In this letter we present such an action and show that it requires a novel type of geometry. This geometry 
does encapsulate NC geometry in its original form when time is absolute (as is the case in Newtonian gravity), but is based
on an underlying symmetry structure, and corresponding set of geometric fields, which goes beyond the Bargmann algebra -
the centrally extended Galilean algebra. 

NC geometry and its recently discovered torsionful version \cite{Christensen:2013rfa,Christensen:2013lma} (referred to
as type I TNC geometry below) has been very useful for studying aspects of field theories with Galilean symmetries. Furthermore, gravity theories for type I TNC geometry (with broken particle number gauge symmetry) have been recently studied as well and shown to correspond to Ho\v rava--Lifshitz gravity, see e.g.  \cite{Hartong:2015zia,Afshar:2015aku}. It has proven difficult to write down actions for type I TNC geometry that preserve  $U(1)$ particle number, though exceptions exist in 2+1 dimensions
\cite{Bergshoeff:2016lwr,Hartong:2016yrf}  but these require an additional field. 

By taking a critical look at Newtonian gravity we will show that an action involving type I TNC geometry is incompatible with the way
in which the  mass source appears in the Poisson equation. This is because in type I TNC geometry mass sources torsion which is not compatible with the notion of absolute time (and hence zero torsion) of Newtonian gravity. The key to identifying the correct geometry lies in carefully considering the
properties of a large speed of light limit of GR, as was recently revisited in \cite{VandenBleeken:2017rij} following
earlier work \cite{Dautcourt:1996pm,Tichy:2011te}. 

We present in this letter a novel type of NC geometry, dubbed type II TNC geometry, which for zero torsion includes the standard (type I) NC geometry used to geometrize Newtonian gravity, and that allows us to formulate an action, in any spacetime dimension $D=d+1$. To this end it is crucial to allow for more general time (lapse) functions than the absolute time of Newtonian gravity. We will show that, while type I TNC geometry follows from gauging the Bargmann algebra \cite{Andringa:2010it,Hartong:2015zia} (see also \cite{Duval:1983pb,DePietri:1994je}), type II TNC geometry follows from a novel non-relativistic symmetry, which turns out to be a non-trivial contraction of the direct sum of the Poincar\'e and  Euclidean algebras in $D=d+1$ dimensions.

The action given in this letter  describes the dynamics of a well-defined truncation of the
non-relativistic limit of GR and has direct physical relevance in a Post-Newtonian regime including the effects of strong gravitational fields, e.g. via gravitational time dilation.  More generally, it can be regarded as providing an off-shell definition of the non-relativistic gravity corner 
of the $Gc\hbar$ Bronstein  cube of physical theories, and as such presents a principle towards constructing a non-relativistic quantum gravity theory. The latter could open up a third road towards (relativistic) quantum gravity, in contradistinction to the 
usually travelled paths approaching it via relativistic quantum field theory or general relativity.

\noindent\textbf{Newton--Cartan Geometry}.

Torsional Newton--Cartan (TNC) geometry consists of a clock-form $\tau_\mu$, a rank-$d$ symmetric tensor   $h_{\mu \nu}$
with signature $(0,1, \ldots , 1)$ and a $U(1)$ connection $m_\mu$. These describe a manifold with a Galilean tangent space structure, geometrizing the Galilean equivalence principle. In Galilean invariant theories the total mass is conserved with the mass current coupling to a $U(1)$ gauge connection $m_\mu$.
The TNC fields transform as tensors under diffeomorphisms ($\xi^\mu$) and exhibit furthermore a set of
local symmetries corresponding to Galilean (or Milne) boosts ($\lambda_\mu$)
and a $U(1)$ gauge transformation  ($\sigma$) associated with mass conservation, 
\begin{eqnarray}
&&\delta\tau_\mu=\mathcal{L}_\xi\tau_\mu\,,\quad\delta h_{\mu \nu} =\mathcal{L}_\xi h_{\mu \nu} +\lambda_\mu \tau_\nu+
\lambda_\nu \tau_\mu \,,\nonumber\\
&&\delta m_\mu=\mathcal{L}_\xi m_\mu+\lambda_\mu+\partial_\mu\sigma\, ,  \label{eq:typeITNC}
\end{eqnarray}
where $\mathcal{L}_\xi$ denotes the Lie derivative along $\xi^\mu$.  The geometric tensors $v^\mu$ and $h^{\mu \nu}$ 
are defined by inverting $ -\tau_\mu\tau_\nu+h_{\mu\nu}$ to  $-v^\mu v^\nu+h^{\mu\nu}$, with the property that $ \tau_\mu h^{\mu\nu}=0$
and  $v^\mu \tau_\mu = -1$.  
The Galilean boost parameters satisfy  $v^\mu\lambda_\mu=0$.  The analogue of the absolute value of the determinant of the pseudo-Riemannian
metric, which for TNC we denote by $e^2$ , is given by  minus the determinant of the matrix $-\tau_\mu\tau_\nu+h_{\mu\nu} $. 
Three useful tensors that are invariant under local Galilean boosts (and rotations)
are: $\hat{v}^\mu\equiv v^\mu-h^{\mu\nu}m_\nu$, $\bar{h}_{\mu\nu}\equiv h_{\mu\nu}-2\tau_{(\mu}m_{\nu)}$ and $\tilde{\Phi} \equiv -v^{\mu}m_{\mu}+\frac{1}{2}h^{\mu\nu}m_{\mu}m_{\nu}$. We also record the completeness relation $-\hat{v}^{\mu}\tau_{\nu}+h^{\mu\lambda}\bar{h}_{\lambda\nu}=\delta_{\nu}^{\mu}$.

We will choose the following affine connection to perform covariant differentiation  
\cite{Bekaert:2014bwa,Jensen:2014aia,Hartong:2014oma,Festuccia:2016awg}
\begin{equation}
\bar{\Gamma}_{\mu\nu}^{\lambda}\equiv-\hat{v}^{\lambda}\partial_{\mu}\tau_{\nu}+\frac{1}{2}h^{\lambda\sigma}\left(\partial_{\mu}\bar{h}_{\nu\sigma}+\partial_{\nu}\bar{h}_{\mu\sigma}-\partial_{\sigma}\bar{h}_{\mu\nu}\right)\,.\label{eq:Special TNC connection}
\end{equation}
This is a metric compatible connection, i.e. $\bar\nabla_\mu\tau_\nu=0=\bar\nabla_\mu h^{\nu\rho}$.
Note that this connection is not invariant under the local $U(1)$ transformation with parameter $\sigma$. In TNC geometry we cannot make  the local Galilean boost and local $U(1)$ symmetries manifest at the same time. 
We also note that this connection has torsion because $\bar{\Gamma}_{[\mu\nu]}^{\lambda}=-\hat{v}^{\lambda}\partial_{[\mu}\tau_{\nu]}$. When the clock 1-form $\tau_\mu$ obeys $h^{\mu\rho}h^{\nu\sigma}\left(\partial_\mu\tau_\nu-\partial_\nu\tau_\mu\right)=0$ we call the torsion twistless and the resulting geometry is called twistless torsional Newton--Cartan (TTNC) geometry \cite{Christensen:2013rfa,Christensen:2013lma,Hartong:2015zia}.  In this work we will assume throughout that $\tau_\mu$ is twistless implying that $\tau_\mu$ obeys the Frobenius integrability condition $\tau_{[\mu}\partial_\nu\tau_{\rho]}=0$ so that $\tau_\mu$ is hypersurface orthogonal. 
Thus in this case the spacetime allows a foliation in terms of equal-time slices.

A useful property of the connection \eqref{eq:Special TNC connection} is $\bar\Gamma^\rho_{\rho\mu}=e^{-1}\partial_\mu e-a_\mu$, where we defined the torsion vector $a_\mu\equiv\mathcal{L}_{\hat v}\tau_\mu$ and $e$ has been defined above. This 
implies $(\bar\nabla_\mu+a_\mu) X^\mu=e^{-1}\partial_\mu(eX^\mu)$.
We define the associated Riemann tensor as usual via
\begin{equation}
\left[\bar\nabla_\mu,\bar\nabla_\nu\right]X_\sigma  =  \bar R_{\mu\nu\sigma}{}^\rho X_\rho-2\bar\Gamma^\rho_{[\mu\nu]}\bar\nabla_\rho X_\sigma\,.
\end{equation}
Further, we define the Ricci tensor as $\bar R_{\mu\nu}=\bar R_{\mu\rho\nu}{}^\rho$. Due to the presence of torsion one can show using the Bianchi identity for $\bar R_{[\mu\nu\sigma]}{}^\rho$, 
that the antisymmetric part of the Ricci tensor is nonzero and equal to
\begin{equation}\label{eq:ASpartRicci}
2\bar R_{[\mu\nu]}=(\tau_\mu a_\nu-\tau_\nu a_\mu)\bar\nabla_\rho\hat v^\rho+\hat v^\rho(\tau_\mu\bar\nabla_\nu a_\rho-\tau_\nu\bar\nabla_\mu a_\rho)\,.
\end{equation}
The above reviewed standard TNC geometry is referred to as type I TNC geometry below.

Finally, we note that a convenient way to think of  type I TNC geometry is via the process of null uplift \cite{Julia:1994bs}, which will
be instrumental below in showing that this geometry cannot correctly describe Newtonian gravity. 
Any TNC geometry can be written as a Lorentzian geometry with a null isometry in one dimension higher. Parameterizing the null isometry with $u$ we can write the Lorentzian metric $\hat g_{MN}$ as
\begin{equation}
\label{eq:nulluplift} 
\hat g_{MN}dx^Mdx^N=2\tau_\mu dx^\mu\left(du-m_\nu dx^\nu\right)+h_{\mu\nu}dx^\mu dx^\nu\,,
\end{equation}
where $x^M=(u,x^\mu)$. The null Killing vector is $\partial_u$. The inverse metric is: $\hat g^{uu}=2\tilde\Phi$, $\hat g^{\mu u}=-\hat v^\mu$ and $\hat g^{\mu\nu}=h^{\mu\nu}$.  At the level of symmetries, the null reduction means that the Bargmann algebra is a subalgebra of Poincar\'e in one dimension higher. 
Alternatively, it can be obtained
by an In\"on\"u-Wigner contraction of the product of the  Poincar\'e algebra (in the same dimension) times a $U(1)$.


\noindent\textbf{A critical look at Newton--Cartan gravity}.

Type I Newton--Cartan geometry was initially invented to describe Newtonian gravity in a coordinate independent manner. The equations of motion that covariantize the Poisson equation of Newtonian gravity are 
\begin{equation}\label{eq:Newt}
\bar R_{\mu\nu}=8\pi G \frac{d-2}{d-1} \rho\tau_\mu\tau_\nu \quad , \quad  \partial_\mu\tau_\nu-\partial_\nu\tau_\mu=0 \, , 
\end{equation}
where $\rho$ is the mass density. On flat spacetime in Cartesian coordinates $\tau=dt$ and $h_{\mu\nu}dx^\mu dx^\nu=dx^idx^i$ with $m=\Phi dt$, this simply reduces to
\begin{equation}
 \partial_i\partial_i\Phi=8\pi G \frac{d-2}{d-1} \rho \, .
 \end{equation} 
Both sides of \eqref{eq:Newt} are formulated in terms of NC objects and are invariant under all type I NC gauge symmetries for $d\tau=0$. Yet, the coupling of mass to the geometry is not what one would expect from a theory with local Bargmann $U(1)$ invariance. The gauge field $m_\mu$ couples to the conserved mass current $J^\mu$, so any type I invariant action leads to an equation of the form $R^\mu=J^\mu$ where the left hand side is a geometrical object formed from the type I TNC fields obeying the Bianchi identity $\partial_\mu\left(eR^\mu\right)=0$.

Using null uplift \eqref{eq:nulluplift}, the equation $R^\mu=J^\mu$ reads $\hat G^\mu{}_u=8\pi G\hat T^\mu{}_u$
where $\hat G^M{}_N$ and $\hat T^M{}_N$ are the higher dimensional Einstein and energy-momentum tensors. From the Bianchi identity for $\hat G^M{}_N$ it follows that $\hat G^\mu{}_u$ is identically conserved. Contracting with $\tau_\mu$ we see that mass sources $\tau\wedge d\tau\neq 0$ since $\hat G_{uu}= 8\pi G \hat T_{uu}\equiv 8\pi G \rho$ 
with $\hat G_{uu}=\frac{1}{4}\left[h^{\mu\rho}h^{\nu\sigma}(\partial_\mu\tau_\nu-\partial_\nu\tau_\mu)(\partial_\rho\tau_\sigma-\partial_\sigma\tau_\rho)\right]$. This conflicts Newtonian gravity since in that case the notion of mass is compatible with $d\tau=0$, i.e. $\rho$ in Newton's law is not a Bargmann mass.

Newtonian gravity is obtained from a non-relativistic limit of GR but we have just shown that this limit cannot be
type I TNC geometry. This begs the question what kind of geometry one should employ. The answer comes from studying the large speed of light limit of GR, i.e. the $1/c$ expansion of \cite{Dautcourt:1996pm,VandenBleeken:2017rij}. We will show that this leads to a different notion of Newton--Cartan geometry which we call type II Newton--Cartan geometry, and that this allows for an off-shell formulation of Newtonian gravity.

\noindent\textbf{$\bf{1/c}$ Expansion and type II TNC geometry}.

In a $1/c$ expansion the pseudo-Riemannian metric and its inverse are expanded as \cite{VandenBleeken:2017rij}
\begin{eqnarray}
\label{eq:gexp}
g_{\mu\nu} & = & -c^2\tau_\mu\tau_\nu+\bar h_{\mu\nu}+\frac{1}{c^2}\left(2\tau_{(\mu}\hat B_{\nu)}-\bar h_{\mu\rho}\bar h_{\nu\sigma}\hat\beta^{\rho\sigma}\right)\nonumber\\
&&+{\mathcal{O}}(c^{-4})\,,\\
g^{\mu\nu} & = & h^{\mu\nu}-\frac{1}{c^2}\hat v^\mu\hat v^\nu+\frac{1}{c^2}\hat\beta^{\mu\nu}+{\mathcal{O}}(c^{-4})\,.
\end{eqnarray}
where we note that
the 1-form $\hat B_\mu$ will play no role in what follows. It is convenient to define $\Phi_{\mu\nu}$ via the relation $\hat\beta^{\mu\nu}=h^{\mu\rho}h^{\nu\sigma}\Phi_{\rho\sigma}$. 

Using the corresponding $1/c$ expansion for the vielbeins \cite{Hansen:2019svu,Hansen:2020pqs} it follows that 
the fields $\tau_\mu$ and $h_{\mu \nu}$  appearing above, transform as in type I TNC geometry (see \eqref{eq:typeITNC}). 
In addition the fields $m_\mu$ and $\Phi_{\mu\nu}$ transform as
\begin{eqnarray}
\label{eq:torsionalU1}
\delta && m_\mu =  \mathcal{L}_\xi m_\mu +  \lambda_\mu +  ( \partial_\mu -a_\mu) \lambda + \tau_\mu h^{\rho \nu} a_\rho \zeta_\nu \\
&& 
\delta\Phi_{\mu\nu} = \mathcal{L}_\xi \Phi_{\mu\nu} -2\lambda\bar K_{\mu\nu}+\bar\nabla_\mu\zeta_\nu+\bar\nabla_\nu\zeta_\mu\,, \nn
\end{eqnarray}
where we defined $\lambda$ and $\zeta_\nu$ through the equation 
 $\zeta^\mu=-\hat v^\mu\lambda+h^{\mu\nu}\zeta_\nu$, $\bar K_{\mu\nu} \equiv - \frac{1}{2}\mathcal{L}_{\hat v}\bar h_{\mu\nu}$ is the extrinsic
 curvature tensor and we recall that $a_\mu\equiv\mathcal{L}_{\hat v}\tau_\mu$ is the torsion vector. 
  These important extra symmetries follow from expanding relativistic diffeomorphism  $\Xi^\mu=\xi^\mu+\frac{1}{c^2}\zeta^\mu+..$, 
 so that $\xi^\mu$ parameterizes non-relativistic diffeomorphisms and $\zeta^\mu$ the extra symmetries above. 
 
 We will refer to the $\lambda$ transformation in \eqref{eq:torsionalU1} as a torsional $U(1)$ transformation due to the presence of the torsion vector $a_\mu$.  One notices that for $d \tau =0$ the transformation of  $m_\mu$ above reduces to the one in  \eqref{eq:typeITNC}, since the
 $\zeta_\nu$ part vanishes in that case while 
  the torsional $U(1)$  takes the same form as the $U(1)$ transformation in  \eqref{eq:typeITNC}. 
   However, the gauge field $m_\mu$ in type II TNC geometry  is quite different from its type I cousin. In particular, we will show in \cite{Hansen:2020pqs}
that in type II TNC geometry $m_\mu$  couples to the energy current as opposed to type I where it couples to the mass current.

Deferring details to \cite{Hansen:2020pqs} we remark that the transformations of the type II TNC geometry introduced above can
be obtained by gauging a novel non-relativistic algebra of dimension $(d+1)(d+2)$, 
spanned by the generators $\{ H, P_a, G_a, J_{ab} \}$ of the (massless) Galilean algebra
augmented with the set  $\{ N, T_a, B_a, S_{ab} \}$, with non-zero commutators: 
\begin{eqnarray}
&&\left[H\,,G_a\right]=P_a\,,\quad \left[P_a\,,G_b\right]=N\delta_{ab}\,,\quad\left[N\,,G_a\right]=T_a\,,\nonumber\\
&&\left[H\,,B_a\right]=T_a\,,\quad\left[S_{ab}\,,P_c\right]=2\delta_{c[a}T_{b]}\,,\quad\left[G_a\,,G_b\right]=-S_{ab}\,,\nonumber\\
&&\left[S_{ab}\,,G_c\right]=2\delta_{c[a}B_{b]}\,,\quad\left[J_{ab}\,,J_{cd}\right]=4\delta_{[a[d}J_{c]b]}\,,\nonumber\\
&&\left[J_{ab}\,,X_c\right]=2\delta_{c[a}X_{b]}\,,\quad\left[J_{ab}\,,S_{cd}\right]=4\delta_{[a[d}S_{c]b]}\,.
\end{eqnarray}
where $X_a\in\{P_a,T_a,G_a,B_a\}$. The first line differs from the Bargmann algebra because $N$ is not central. Interestingly, this algebra can be obtained from a contraction of the direct sum of the Poincar\'e and Euclidean algebras in $d+1$ dimensions \cite{Note1}
and underlies Newtonian gravity in the same way that the Poincar\'e algebra underlies GR.

\noindent\textbf{Off-shell Newtonian gravity}.

We now construct a Lagrangian depending on $\tau_\mu$, $h_{\mu\nu}$, $m_\mu$, $\Phi_{\mu\nu}$ that is invariant under the above gauge transformations. The unique two-derivative result is
\begin{eqnarray}
\mathcal{L} & = & -\frac{1}{16 \pi G }e\left[\hat v^\mu\hat v^\nu\bar R_{\mu\nu}-\tilde\Phi h^{\mu\nu}\bar R_{\mu\nu}\right.\nonumber\\
&&\left.-\Phi_{\mu\nu}h^{\mu\rho}h^{\nu\sigma}\left(\bar R_{\rho\sigma}-a_\rho a_\sigma-\bar\nabla_{\rho}a_{\sigma}\right)\right.\nonumber\\
&&\left.+\frac{1}{2}\Phi_{\mu\nu}h^{\mu\nu}\left[h^{\rho\sigma}\bar R_{\rho\sigma}-2e^{-1}\partial_\rho\left(eh^{\rho\sigma}a_\sigma\right)\right]\right]\,, 
\end{eqnarray}
where $e$ is the integration measure and we have omitted a possible cosmological constant term $e\Lambda$. 
The Lagrangian is obtained by starting with the (necessary) kinetic term $\hat v^\mu\hat v^\nu\bar R_{\mu\nu}$ and subsequently adding terms such that
the entire expression  is invariant under the torsional $U(1)$ transformation as well the $\zeta_\mu$ transformation \eqref{eq:torsionalU1}. This 
invariance follows from the Bianchi identities
\begin{eqnarray}
0 & = & e^{-1}\partial_\rho\left(e\left[h^{\rho\nu}\hat v^\mu\bar R_{\mu\nu}-\frac{1}{2}\hat v^\rho h^{\mu\nu}\bar R_{\mu\nu}\right]\right)\nonumber\\
&&+h^{\mu\rho}h^{\nu\sigma}\bar K_{\rho\sigma}\bar R_{\mu\nu}-\frac{1}{2}h^{\rho\sigma}\bar K_{\rho\sigma}h^{\mu\nu}\bar R_{\mu\nu}\,,\\
0 &= & h^{\mu\nu}h^{\rho\sigma}\bar\nabla_\mu\bar R_{\nu\rho}
-\frac{1}{2}h^{\mu\sigma}h^{\nu\rho}\bar\nabla_\mu\bar R_{\nu\rho}\,,
\end{eqnarray}
which can be derived from  $\bar\nabla_{[\lambda}\bar R_{\mu\nu]\sigma}{}^\kappa=0$.  

Since we work with off-shell TTNC geometries we need to add the Lagrange multiplier term $\mathcal{L}_{LM}=e\zeta^{\mu\nu}(\partial_\mu\tau_\nu-\partial_\nu\tau_\mu)$ to the Lagrangian where $\zeta^{\mu\nu}=-\zeta^{\nu\mu}$ obeying $\tau_\mu\zeta^{\mu\nu}=0$ so that it only imposes $\tau\wedge d\tau=0$ but not $d\tau=0$ \cite{Note2}. 
If we were to drop the condition $\tau_\mu \zeta^{\mu\nu}=0$ so that $\mathcal{L}_{LM}$ enforces $d\tau=0$, the field $\zeta^{\mu\nu}$ would not decouple from the equations of motion. 
This is what happens in the 3D Chern-Simons actions for extended Bargmann algebras \cite{Bergshoeff:2016lwr,Hartong:2016yrf} where $\zeta^{\mu\nu}=\epsilon^{\mu\nu\rho}\zeta_\rho$ with $\zeta_\rho$ associated to the central extension of the 3D Bargmann algebra. 


We are going to compute the equations of motion by varying $\tilde\Phi$, $\hat v^\mu$, $\Phi_{\mu\nu}$ and $h^{\mu\nu}$. Let us define 
\begin{equation}
\delta\mathcal{L}=- \frac{e}{8\pi G} \left(E_{\tilde\Phi}\delta\tilde\Phi-E_\mu\delta\hat v^\mu+\frac{1}{2}E^h_{\mu\nu}\delta h^{\mu\nu}+\frac{1}{2}E^{\mu\nu}\delta\Phi_{\mu\nu}\right)\,,
\end{equation}
where $E_{\tilde\Phi} = -\frac{1}{2}h^{\mu\nu}\bar R_{\mu\nu}$ and
\begin{eqnarray}
E^{\mu\nu} &=& -h^{\mu\rho}h^{\nu\sigma}\left(\bar R_{\rho\sigma}-a_\rho a_\sigma-\bar\nabla_{\rho}a_{\sigma}\right)\nonumber\\
&&+\frac{1}{2}h^{\mu\nu}\left(h^{\rho\sigma}\bar R_{\rho\sigma}-2e^{-1}\partial_\rho\left(eh^{\rho\sigma}a_\sigma\right)\right)\,.
\end{eqnarray}
The variations with respect to $P^\rho_\mu\delta \hat v^\mu$ with $P^\rho_\mu$ the spatial projector $P_\mu^\rho\equiv\delta_\mu^\rho+\hat v^\rho\tau_\mu$ gives
\vskip -.5cm
\begin{equation}
h^{\rho\mu}E_\mu = -h^{\rho \mu} \hat v^\nu \bar R_{\mu\nu} \ . 
\end{equation}
The remaining variations are $\tau_\mu\delta\hat v^\mu$ and $P^\alpha_\mu P^\beta_\nu\delta h^{\mu\nu}$. Defining $E_h^{\alpha\beta}\equiv h^{\mu\alpha}h^{\nu\beta}E_{\mu\nu}^h$ we find
\vskip -0.2cm
\begin{widetext}
\begin{eqnarray}
-2\hat v^\mu E_\mu & = &-2\tilde\Phi E_{\tilde\Phi}-\Phi_{\mu\nu}E^{\mu\nu}-h^{\mu\nu}\Phi_{\mu\nu}e^{-1}\partial_\rho\left(eh^{\rho\sigma}a_\sigma\right)+h^{\mu\rho}h^{\nu\sigma}\Phi_{\mu\nu}\left(\bar\nabla_\rho a_\sigma+a_\rho a_\sigma\right)+\left(h^{\rho\sigma}\bar K_{\rho\sigma}\right)^2\nonumber\\
&&-h^{\rho\sigma}h^{\kappa\lambda}\bar K_{\rho\kappa}\bar K_{\sigma\lambda}+\bar\nabla_\mu\left[h^{\mu\rho}h^{\nu\sigma}\left(\bar\nabla_\rho\Phi_{\nu\sigma}-\bar\nabla_\nu\Phi_{\rho\sigma}\right)\right]\,,\\
E_h^{\alpha\beta} & = & 
\left(h^{\mu\alpha}h^{\nu\beta}\Phi_{\mu\nu}-\frac{1}{2}h^{\alpha\beta}h^{\mu\nu}\Phi_{\mu\nu}\right)\left(e^{-1}\partial_\rho\left(eh^{\rho\sigma}a_\sigma\right)+E_{\tilde\Phi}\right)
-\frac{1}{2}h^{\alpha\beta}\Phi_{\mu\nu}E^{\mu\nu}+h^{\mu\alpha}\Phi_{\mu\rho}E^{\rho\beta}+h^{\mu\beta}\Phi_{\mu\rho}E^{\rho\alpha}\nonumber\\
&&-\frac{1}{2}h^{\rho\sigma}\Phi_{\rho\sigma}E^{\alpha\beta}+\tilde\Phi E^{\alpha\beta}\nonumber-\frac{1}{2}h^{\alpha\beta}\left[\left(h^{\mu\nu}\bar K_{\mu\nu}\right)^2-h^{\mu\rho}h^{\nu\sigma}\bar K_{\mu\nu}\bar K_{\rho\sigma}\right]+\bar\nabla_\rho\left[\hat v^\rho h^{\mu\alpha}h^{\nu\beta}\bar K_{\mu\nu}-\hat v^\rho h^{\alpha\beta}h^{\mu\nu}\bar K_{\mu\nu}\right]\nonumber\\
&&+h^{\mu\alpha}h^{\nu\beta}\bar\nabla_\mu\partial_\nu\tilde\Phi+h^{\mu\alpha}h^{\nu\beta}\left(a_\mu\partial_\nu\tilde\Phi+a_\nu\partial_\mu\tilde\Phi\right)-h^{\alpha\beta}h^{\mu\nu}\bar\nabla_\mu\partial_\nu\tilde\Phi-2h^{\alpha\beta}h^{\mu\nu}a_\mu\partial_\nu\tilde\Phi\nonumber\\
&&-\frac{1}{2}h^{\alpha\beta}h^{\mu\nu}h^{\rho\sigma}\left(\bar\nabla_\mu+a_\mu\right)\left(\bar\nabla_\rho+a_\rho\right)\Phi_{\nu\sigma}+h^{\mu\alpha}h^{\nu\beta}h^{\rho\sigma}\left(\bar\nabla_\rho+a_\rho\right)\left(\bar\nabla_{(\mu}\Phi_{\nu)\sigma}-\frac{1}{2}\bar\nabla_\sigma\Phi_{\mu\nu}\right)\nonumber\\
&&+\frac{1}{2}h^{\alpha\beta}h^{\mu\nu}h^{\rho\sigma}\left(\bar\nabla_\mu+a_\mu\right)\bar\nabla_\nu\Phi_{\rho\sigma}-\frac{1}{2}h^{\mu\alpha}h^{\nu\beta}h^{\rho\sigma}\bar\nabla_\mu\bar\nabla_\nu\Phi_{\rho\sigma}\,.
\end{eqnarray}
\end{widetext}
We only need to consider the variation $P^\alpha_\mu P^\beta_\nu\delta h^{\mu\nu}$ because we are only interested in the spatial projection of $\Phi_{\mu\nu}$. By taking the trace of $E_h^{\alpha\beta}$ and using $\hat v^\mu E_\mu$ we find
\begin{eqnarray}
h^{\mu\nu}E_{\mu\nu}^h & = & -(d-2)\hat v^\mu E_\mu+\Phi_{\mu\nu}E^{\mu\nu}-(d-1) [ \hat v^\mu\hat v^\nu\bar R_{\mu\nu} \nonumber\\
&&\hspace{-1.5cm} - \left(\bar\nabla_\mu+a_\mu\right)\left(h^{\mu\nu}a_\nu\left(\tilde\Phi+\frac{1}{2}h^{\rho\sigma}\Phi_{\rho\sigma}\right)-h^{\mu\nu}h^{\rho\sigma}a_\rho\Phi_{\nu\sigma}\right) ] \,,\nonumber\\
&& \label{eq:traceeq}
\end{eqnarray}
\vskip -.7cm
\noindent where we used the identity
\begin{eqnarray}
\hspace{-.2cm}\hat v^\mu\hat v^\nu\bar R_{\mu\nu} & = & \left(h^{\mu\nu}\bar K_{\mu\nu}\right)^2-h^{\mu\rho}h^{\nu\sigma}\bar K_{\mu\nu}\bar K_{\rho\sigma}+3h^{\mu\nu}a_\mu\partial_\nu\tilde\Phi\nonumber\\
\hspace{-.2cm}&&\hspace{-1.5cm}+\bar\nabla_\mu\left(\hat v^\mu h^{\nu\rho}\bar K_{\nu\rho}+h^{\mu\nu}\partial_\nu\tilde\Phi\right)+2\tilde\Phi e^{-1}\partial_\mu\left(eh^{\mu\nu}a_\nu\right)\,.
\end{eqnarray}
Note that for $d \tau =0$ the field $\Phi_{\mu \nu}$ decouples. 

It can be shown that these equations agree with \cite{VandenBleeken:2017rij}, where they were  obtained by expanding the Einstein equations in $1/c^2$. However \cite{VandenBleeken:2017rij}
 did not determine the equations of motion for $h^{\alpha\mu}h^{\beta\nu}\Phi_{\mu\nu}$ which we obtain by varying $\hat v^\mu$ and $h^{\mu\nu}$. These equations are essential in order to obtain a closed system of equations for the general case $d\tau\neq 0$. Importantly, we note that our action allows for geometries with strong gravitational fields, and in particular those with $\tau$ not closed allow for non-relativistic gravitational time dilation. 

Given the gravity action with type II TNC gauge invariance we need to understand how matter couples to such a geometry. This will be discussed in \cite{Hansen:2020pqs}, but as remarked before this
coupling will be markedly different than the known couplings of matter to type I TNC geometry
\cite{Jensen:2014aia,Hartong:2014pma,Geracie:2015xfa,Fuini:2015yva,Bergshoeff:2015sic,Festuccia:2016caf}. 
One of the reasons is that while in type I TNC geometry  $m_\mu$ couples to the mass current, in type II it couples to the energy  current. 
This will be further discussed in \cite{Hansen:2020pqs} by carefully studying the $1/c$ limit of the worldline action of a relativistic particle as well
as the known couplings of Poincar\'e invariant field theories to pseudo-Riemannian geometry. 

Here we will consider only the very special case of a static  particle in order to obtain the Poisson equation from an action principle. A static point mass with mass density $\rho$ has a Lagrangian that is simply $\mathcal{L}_m=\alpha e\rho$ with 
 $\alpha= -\frac{d-2}{2}$. 
 Taking the trace of $E^{\mu\nu}$ gives $\bar h_{\mu\nu}E^{\mu\nu}+(d-2)E_{\tilde\Phi}=-(d-1)e^{-1}\partial_\rho\left(eh^{\rho\sigma}a_\sigma\right)$.
Varying $\mathcal{L}+\mathcal{L}_m$ tells us that the left hand side vanishes and hence that $\partial_\mu\left(e h^{\mu\nu}a_\nu\right)=0$. Since $\tau\wedge d\tau=0$ we have that $h^{\mu\rho}h^{\nu\sigma}\left(\partial_\rho a_\sigma-\partial_\sigma a_\rho\right)=0$, so that $h^{\mu\nu}a_\nu=h^{\mu\nu}\partial_\nu F$ for some function $F$. Hence $\partial_\mu\left(e h^{\mu\nu}a_\nu\right)=0$ states that $F$ is a harmonic function on the $d$-dimensional Riemannian geometry of the hypersurface to which $\tau$ is orthogonal. Regularity requires $F$ to be constant and hence that $d\tau=0$, as desired in Newtonian gravity which has absolute time. 
What survives from \eqref{eq:traceeq} is then the equation  $(d-1) \hat v^\mu\hat v^\nu\bar R_{\mu\nu} = -(d-2)\hat v^\mu E_\mu  -
h^{\mu\nu}E_{\mu\nu}^h $. Then, taking into account the matter contribution
to $E_{\mu}$  and $E_{\mu\nu}^h$  in this equation,
for which we use the variation 
$\delta {\cal{L}}_m  = \alpha e \rho (\tau_\mu \delta \hat v^\mu 
- \frac{1}{2} h_{\mu \nu} \delta h^{\mu \nu} ) $,
it follows that  the equations of motion of $\mathcal{L}+\mathcal{L}_m$  with $d\tau=0$  are nothing else but Newton's law \eqref{eq:Newt}. 

\noindent\textbf{Discussion}.
\vskip -.1cm

Among the numerous avenues that one may pursue following our action and corresponding novel geometry 
we mention a few. It would be interesting to: i). examine if there exists a geometric construction that gives type II TNC geometry from some Lorentzian starting point, just like type I follows from null reduction of a Lorentzian metric, ii). perform a Hamiltonian analysis along with
determining the asymptotic symmetries and examining the solution space of the theory, iii). work out how particles, strings and branes probe type II TNC geometry
and see if the equations of motion of the non-relativistic gravity action can be related to consistency conditions of some type of string theory (see  \cite{Harmark:2017rpg,Bergshoeff:2018yvt,Harmark:2018cdl} for non-relativistic strings in the context of type I TNC geometry). Finally, there are undoubtedly also exciting applications in the realm of the AdS/CFT correspondence and generalizations thereof. 

\noindent\textbf{Acknowledgements.}

We thank  Eric Bergshoeff, Dieter van den Bleeken, Shira Chapman,  Lorenzo Di Pietro, Jos\'e Figueroa-O'Farrill, Gerben Oling, Manus Visser 
and  Ziqi Yan, for useful discussions.
The work of DH is supported by the Swiss National Science Foundation through the NCCR SwissMAP.
The work of JH is supported by the Royal Society University Research Fellowship ``'Non-Lorentzian Geometry in Holography'' (grant number UF160197).
The work of NO is supported in part by the project ``Towards a deeper understanding of  black holes with non-relativistic holography'' of the Independent Research Fund Denmark (grant number DFF-6108-00340).  JH and NO thank the Perimeter Institute for hospitality during completion of this work.

\providecommand{\href}[2]{#2}\begingroup\raggedright\endgroup

\end{document}